\documentstyle[12pt,psfig]{article}

\def\thalf{{\textstyle{\frac{1}{2}}}}
\def\tquar{{\textstyle{\frac{1}{4}}}}
\def\tthralf{{\textstyle{\frac{3}{2}}}}

\title{\begin{flushright}
{\normalsize CERN-TH/99-193 \\ NUC-MINN-99/9-T\\
July 1999 \\ Revised January 2000\\}
\end{flushright}
\vspace*{0.1in}
{\bf DYNAMICAL EVOLUTION OF THE SCALAR CONDENSATE IN HEAVY ION COLLISIONS}}
\author{{\bf L\'aszl\'o P. Csernai}$^{1-3}$, {\bf Paul J. Ellis}$^3$\\
{\bf Sangyong Jeon}$^4$, {\bf Joseph I. Kapusta}$^{3,5}$ \vspace*{0.1in} \\
 $^1${\it Department of Physics, University of Bergen}\\ \vspace*{0.2in}
 {\it 5007 Bergen, Norway}\\
 $^2${\it KFKI Research Institute for Particle \& Nuclear Physics}\\
 \vspace*{0.2in}     {\it P.O. Box 49, 1525 Budapest, Hungary}\\
 $^3${\it School of Physics and Astronomy, University of Minnesota} \\
 \vspace*{0.2in}
 {\it Minneapolis, MN 55455, USA}\\
 $^4${\it Nuclear Science Division, Lawrence Berkeley National Laboratory} \\
 \vspace*{0.2in}
 {\it Berkeley, CA 94720, USA}\\
   $^5${\it TH Division, CERN}\\
   {\it CH-1211 Geneva 23, Switzerland}}

\date{}

\begin{document}

\maketitle
\begin{abstract}

We derive the effective coarse-grained field equation for the scalar condensate
of the linear sigma model in a simple and straightforward manner using linear
response theory.  The dissipative coefficient is calculated at tree level 
on the basis of the physical processes of sigma-meson decay and of thermal 
sigma-mesons and pions knocking sigma-mesons out of the condensate.  The field
equation is solved for hot matter undergoing either one or three dimensional
expansion and cooling in the aftermath of a high energy nuclear collision.  The
results show that the time constant for returning the scalar condensate 
to thermal equilibrium is of order 2 fm/c.

\end{abstract}

\section{Introduction}

Scalar condensates often appear in quantum field theories when a symmetry 
is spontaneously broken.  Prominent examples include the Higgs condensate and
the chiral condensate.  The equilibrium behavior of these condensates as a
function of temperature and density has been extensively studied in the context
of cosmology and heavy ion collisions.  However, much remains to be learned
about how these condensates really evolve in out-of-equilibrium systems.

In this paper, we study the dynamical evolution of the chiral condensate in the
$O(N)$ linear sigma model and apply it to the expanding matter in a high energy
heavy ion collision.  We recall that the sigma field $\sigma$ represents the
quark condensate $\bar{q}q$ in the sense that they both have the same
transformation properties for $N=4$, corresponding to two flavors of massless
quarks.  At high temperatures, quarks and gluons exist in a deconfined, chirally
symmetric phase.  At some critical temperature of order 150 MeV a transition to
a hadronic phase occurs.  In this confined and symmetry-broken phase the quark
condensate is nonzero.  In a collision between large nuclei the beam energy must
be high enough so that the matter reaches at least approximate thermal
equilibrium at a temperature greater than the critical temperature.  We derive
and solve an effective coarse-grained equation of motion for the chiral
condensate, or mean sigma field, starting at the critical temperature and
continuing down to the temperature where the system loses its ability to
maintain local equilibrium and freezes out.

The equation of motion approach we espouse here was used by Linde in his
pioneering work on phase transitions in relativistic quantum field theory
\cite{Linde}, although he did not use this technique to analyze dissipative
processes or fields out of equilibrium.  Our derivation of a coarse-grained
field equation is based on linear response theory \cite{FW}, where the 
connection between finite temperature averages and the evolution of an 
observable in real physical time is clear and straightforward.  There is less 
clarity, in our opinion, with the Influence Functional \cite{IF} and closely 
related Closed Time Path \cite{CTP} methods as used by many others in this 
context \cite{GR,Boya,GM,Rischke,GL}.  Full nonlinear response theory is 
certainly equivalent to both of those methods, as indeed it must be since they 
all describe the same physics.  However, in most cases practical calculations 
can only be performed when the deviation from thermal equilibrium is small.  
This is true of the Influence Functional and Closed Time Path methods as well as 
response theory.  Still we are able to make several improvements to the most 
recent treatment \cite{Rischke} of the linear sigma model.  First, the 
complications associated with doubling of the field variables do not arise, nor 
those arising from paths in the complex time plane.  Second, we make a direct 
connection with physical response functions.  It is clear that these response 
functions should be evaluated in the fully interacting equilibrium ensemble.  
Here we shall estimate the relevant response functions by including the physical 
processes of sigma formation and decay via two pions, and the scattering of 
thermal pions and sigma mesons from sigma mesons in the condensate.  To our 
knowledge, the latter processes have not been studied before in this or any 
related context.  The amplitudes for these processes are computed in tree 
approximation.

In section 2 we derive the fundamental equation.  In section 3 we calculate
decay and scattering contributions to the dissipative term in the coarse-grained
field equation for the condensate.  In section 4 we solve the equation of motion
in the hadronic phase of a high energy heavy ion collision.  Conclusions, and
extensions of this paper, are discussed in section 5.

\section{The Fundamental Equation}

The Lagrangian of the linear $O(N)$ sigma model involves the sigma field
and a vector of $(N-1)$ pion fields, $\mbox{\boldmath $\pi$}$:
\begin{equation}
{\cal L} = \thalf\left(\partial_{\mu}\sigma\right)^2 +
\thalf\left(\partial_{\mu}\mbox{\boldmath $\pi$}\right)^2
- \tquar\lambda\left(\sigma^2+\mbox{\boldmath $\pi$}^2 -
f_{\pi}^2\right)^2\;,
\end{equation}
where $\lambda$ is a positive coupling constant and $f_{\pi}$ is the pion decay
constant.  In the vacuum the symmetry is spontaneously broken: the sigma field
acquires a vacuum expectation value of $\langle 0 | \, \sigma | 0 \rangle =
f_{\pi}$, excitations of the sigma field have mass $m_{\sigma}^2 = 2 \lambda
f_{\pi}^2$, and the pion is a Goldstone boson since we neglect explicit
chiral symmetry breaking here.  The symmetry is restored by a
second order phase transition at the critical temperature $T_c^2 = 12
f_{\pi}^2/(N+2)$.

When one is interested in what happens at a fixed temperature at and below 
$T_c$ the sigma field is usually expressed as
\begin{equation}
\sigma({\bf x},t) = v + \sigma'({\bf x},t)\;,
\end{equation}
where the thermal average of the sigma field at temperature $T$ is $\langle
\sigma \rangle_{eq} = v$ so that $\langle \sigma' \rangle_{eq} = 0$.  The
equation of
motion for the fluctuating component $\sigma'$ is
\begin{equation}
\ddot{{\sigma}}' - \nabla^2 \sigma' = \lambda f_{\pi}^2 \left(v+\sigma' \right)
- \lambda \left( v+\sigma' \right)^3 - \lambda \left( v+\sigma' \right)
\mbox{\boldmath $\pi$}^2 \, .
\end{equation}
We now allow the sigma field to be slightly out of equilibrium, and write
instead:
\begin{equation}
\sigma({\bf x},t) = v + \sigma_s({\bf x},t) + \sigma_f({\bf x},t) \, .
\end{equation}
Here $\langle \sigma \rangle = v + \sigma_s$, where $v$ denotes the equilibrium
value as before, but the deviation has been
split into two pieces: a slow part $\sigma_s$, whose average is nonzero, and a
fast part $\sigma_f$, whose average is zero.
(Primes have been dropped for clarity of presentation.)  The notation
$\langle \sigma \rangle$ denotes averaging over the space-time volume
chosen for coarse-graining and thus the precise division
between the fast and slow parts of the field depends on the choice that is
made. For example, one may include in the slow part only those
Fourier components with wavenumber below some cutoff value \cite{BMS,LM}, or one
may average the field fluctuations over some coarse-graining time \cite{CJK}.
We shall not delve into any details here, but only remark that if the results
depend strongly on the coarse-graining technique then the procedure is not very
useful in the given context.  In any case, the slow part represents occupation
of the low momentum modes by a large number of particles and so may be treated
as a slowly varying classical field.   The fast part is a fully quantum field.
The equilibrium ensemble average of an operator $O$ is denoted by $\langle O
\rangle_{eq}$ and is characterized by the temperature $T$ and the full
Hamiltonian determined from the linear sigma model Lagrangian in the absence of 
$\sigma_s$.  When this equilibrium ensemble is perturbed by the presence of 
the field $\sigma_s$ the Hamiltonian is modified and the resulting
(nonequilibrium) ensemble average is denoted by $\langle O \rangle$.

It should be noted at this point that we have not allowed for a nonzero ensemble
average of the pion field.  Such a nonzero average is usually referred to as a
disoriented chiral condensate, or DCC.  One could certainly allow for a DCC and
carry through the following computations in a straightforward way.  However
in thermal equilibrium a DCC never develops (although it can arise from
thermal fluctuations in a small system, but even then the probability is
small \cite{csernai}).  This is in contrast to the scalar
condensate, whose value is zero above the critical temperature and becomes
nonzero below it.  (A very small pion mass, too small even to affect the
equation of state or correlation functions at the temperatures of interest,
will still tilt the system towards a unique vacuum.) Thus we choose to focus
on the behavior of the $\sigma$ field.

Let us average the sigma field equation over time and length scales large
compared to the scales characterizing the quantum fluctuations of the fields
$\sigma_f$ and $\mbox{\boldmath $\pi$}$ but short compared to the scales
typifying $\sigma_s$.  It is an assumption that such a separation exists, but in
any given situation it can be verified or refuted {\it a posteriori}.  Since
$\langle \sigma_f \rangle = 0$ we obtain
\begin{equation}
\ddot{{\sigma}}_s - \nabla^2 \sigma_s = \lambda \left(v+\sigma_s \right)
\left[f_{\pi}^2 - \left(v+\sigma_s \right)^2 - 3\langle \sigma_f^2 \rangle -
\langle \mbox{\boldmath $\pi$}^2 \rangle \right] - \lambda \langle \sigma_f^3
\rangle \, .
\end{equation}
The full ensemble averages are
\begin{eqnarray}
\langle \sigma_f^2 \rangle &=& \langle \sigma_f^2 \rangle_{eq} + \delta \langle
\sigma_f^2 \rangle \\
\langle \mbox{\boldmath $\pi$}^2 \rangle &=& \langle \mbox{\boldmath $\pi$}^2
\rangle_{eq} + \delta \langle \mbox{\boldmath $\pi$}^2 \rangle \\
\langle \sigma_f^3 \rangle &=& \langle \sigma_f^3 \rangle_{eq} + \delta \langle
\sigma_f^3 \rangle\;,
\end{eqnarray}
where the deviations are caused by $\sigma_s$ and are generally proportional to
$\sigma_s$ to some positive power.  Equation (5) must be satisfied even when
$\sigma_s$ vanishes.  That determines the equilibrium value of $v$ to be
\begin{equation}
v=0 \;\;\; {\rm if} \;\;\; T > T_c
\end{equation}
or
\begin{equation}
v^2 = f_{\pi}^2 - 3 \langle \sigma_f^2 \rangle_{eq} - \langle \mbox{\boldmath
$\pi$}^2 \rangle_{eq} - \langle \sigma_f^3 \rangle_{eq}/v \;\;\; {\rm if} \;\;\;
T < T_c \, .
\end{equation}
We are interested in the second of these because it represents the low
temperature symmetry-broken phase.  To first approximation in either (i) a 
perturbative expansion in $\lambda$ or (ii) an expansion in $1/N$ \cite{vs}, 
which are the usual approximations for the sigma model, the field fluctuations 
are
\begin{equation}
\langle \sigma_f^2 \rangle_{eq} = \int \frac{d^3p}{(2\pi)^3}
\frac{1}{E_{\sigma}} n_B(E_{\sigma}/T) \, ,
\end{equation}
where $ E_{\sigma} = \sqrt{m_{\sigma}^2+p^2}$ and $n_B(E/T) = 1/\left(
\exp(E/T)-1 \right)$ is the Bose-Einstein distribution.  Note that the sigma
mass is still to be determined.  The pion fluctuations are
\begin{equation}
\langle \mbox{\boldmath $\pi$}^2 \rangle_{eq} = (N-1)\frac{T^2}{12} \, .
\end{equation}
The latter follows from the fact that there are $N-1$ Goldstone bosons.
Corrections to these formulae come from interactions not included in the
effective mass.  The sigma mass is very large at zero temperature and vanishes
at $T_c$.  The term $\langle \sigma_f^3 \rangle_{eq}/v$ is not zero
on account of the cubic self-coupling of the $\sigma$ with coupling coefficient
$\lambda v$.  It does have a finite limit as $v \rightarrow 0$.  However, 
once the approximations (11--12) have been made it is not legitimate to keep 
this term because it is one higher power in $\lambda$ and/or $1/N$, and keeping 
it would violate the $O(N)$ symmetry.  Therefore the limits are:
\begin{eqnarray}
T \rightarrow 0: && v^2 = f_{\pi}^2 - (N-1)T^2/12 \nonumber \\
T \rightarrow T_c: && v^2 = f_{\pi}^2 - (N+2)T^2/12
\end{eqnarray}
where the formula for $T_c$ was given earlier.  It is, of course, consistent
with the above expression.

Using Eqs. (6--8) and (10) in (5), the coarse-grained field equation for
$\sigma_s$
can now
be written as
\begin{equation}
\ddot{{\sigma}}_s - \nabla^2 \sigma_s + 2\lambda v^2 \sigma_s = -\lambda \left[
3v\sigma_s^2 + \sigma_s^3 + \delta \langle \sigma_f^3 \rangle + (v +
\sigma_s)(3\delta\langle \sigma_f^2 \rangle +
\delta \langle \mbox{\boldmath $\pi$}^2 \rangle ) \right] \, .
\end{equation}
Note that here the system is assumed to be at a fixed temperature so that 
$v(T)$ is a constant.  If the temperature is allowed to vary then time 
derivatives of $v$ must be included; see the next section.  Equation (14) shows 
that the sigma mass and the equilibrium value of the scalar
condensate are related by $m_{\sigma}^2 = 2\lambda v^2$.  They are temperature
dependent and determined self-consistently from the formula
\begin{equation}
m_{\sigma}^2 = 2\lambda \left[ f_{\pi}^2 - (N-1)\frac{T^2}{12} -3 \int
\frac{d^3p}{(2\pi)^3} \frac{1}{E_{\sigma}} n_B(E_{\sigma}/T) \right] \, .
\end{equation}
The limits $T \rightarrow 0$ and $T \rightarrow T_c$ are readily obtained from
those of $v$.  An interpolating formula which connects these two limits is:
\begin{equation}
\frac{m_{\sigma}^2}{2\lambda f_{\pi}^2} = \frac{v^2}{f_{\pi}^2} \approx {1 -
T^2/T_c^2
\over
1 - [3/(N+2)]\, (T^2/T_c^2)\, (1 - T^2/T_c^2) } \, .
\end{equation}
This is useful when studying solutions of the equation of motion numerically.

The deviations in the fluctuations due to the presence of $\sigma_s$ give rise
to a renormalization of the parameters in the equation of motion of $\sigma_s$
but, more importantly, they lead to dissipation.  Energy can be transferred 
between the field $\sigma_s$ and the fields $\sigma_f$ and
$\mbox{\boldmath $\pi$}$.  These deviations in the fluctuations may be computed 
using linear response theory \cite{FW,mybook} as long as $\sigma_s$ is small.  
Here small means in comparison to either $v$ (most relevant at low temperature) 
or to $\sqrt{\langle \sigma_f^2 \rangle}$ (most relevant at high temperature).  
Technically, linear response theory is an expansion in powers of the Hamiltonian 
coupling the out-of-equilibrium field $\sigma_s$ to the other modes of the 
system, and guarantees that symmetries of the theory are respected.  This piece 
of the Hamiltonian consists of positive powers of $\sigma_s$ and so becomes 
smaller and smaller with decreasing departures from equilibrium.  The coupling 
between the slow modes and the fast modes are determined straightforwardly from 
the potential to be
\begin{equation}
H_{s\pi} = \lambda \left( v \sigma_s + \thalf \sigma_s^2 \right)
\mbox{\boldmath $\pi$}^2
\end{equation}
and
\begin{equation}
H_{sf} = \lambda\left[ \sigma_s^2 + 3v\sigma_s + 3v^2 -f_{\pi}^2
+\tthralf(\sigma_s +2v)\sigma_f + \sigma_f^2 \right] \sigma_s \sigma_f \, .
\end{equation}
In order to apply linear response analysis we need some initial conditions.  In
a nuclear collision, or in the early universe for that matter, it is assumed
that the system reaches a state of thermal equilibrium at some negative time,
that the system expands and cools, and at time $t=0$ the system is at the
critical temperature.  This implies the initial condition $\sigma_s({\bf x},t=0)
= 0$.
(In a finite volume one may wish to consider an ensemble of initial values
chosen from a canonical distribution \cite{csernai}.) From the standard
theory of linear response one immediately deduces that
\begin{eqnarray}
\delta \langle \sigma_f^n(x) \rangle &=& i \lambda\int_0^t dt' \int
d^3x' \left( \left( 3v^2 - f_{\pi}^2 \right) \sigma_s(x') +3v\sigma_s^2(x')
+\sigma_s^3(x')
\right) \left\langle \left[\sigma_f(x'), \sigma_f^n(x) \right]
\right\rangle_{eq}\nonumber \\
&+& i 3\lambda \int_0^t dt' \int
d^3x' \left( v \sigma_s(x') + \thalf\sigma_s^2(x')
\right) \left\langle \left[\sigma_f^2(x'), \sigma_f^n(x) \right]
\right\rangle_{eq} \nonumber \\
&+& i \lambda\int_0^t dt' \int
d^3x' \sigma_s(x') \left\langle \left[\sigma_f^3(x'), \sigma_f^n(x) \right]
\right\rangle_{eq}\;, \label{delsig}
\end{eqnarray}
and
\begin{equation}
\delta \langle \mbox{\boldmath $\pi$}^2(x) \rangle = i \lambda \int_0^t
dt' \int d^3x' \left( v \sigma_s(x') + \thalf\sigma_s^2(x') \right)
\langle \left[\mbox{\boldmath $\pi$}^2(x'),
\mbox{\boldmath $\pi$}^2(x) \right] \rangle_{eq} \, . \label{delpi}
\end{equation}
The response functions are just the commutators of powers of the field
operators evaluated at two different space-time points in the
unperturbed (equilibrium) ensemble.  We should emphasize that the unperturbed
ensemble {\em does} include all interactions among the fast modes and makes no
approximation regarding the strength of these interactions.  Insertion of these
deviations into Eq. (14) represents the fundamental equation of this paper.
For very small departures from equilibrium it is sufficient to keep only those
terms which are linear in $\sigma_s$.  The high temperature, symmetric phase is
obtained by setting $v =0$.

The time-delayed response of the fast modes to the slow one evident in eqs. 
(19-20) has two effects: The
sigma mass and self-interactions are modified, and dissipation occurs.  These
effects may be seen by expanding the slow field $\sigma_s({\bf x}',t')$ in a
Taylor series about the point $({\bf x},t)$ in
Eqs. (\ref{delsig}--\ref{delpi}) (although such an
expansion is not required).  Terms with no derivative or an even number of
derivatives either renormalize existing terms in the equation of motion or add
new nondissipative ones, such as $\sigma_s \nabla^2 \sigma_s$ and $\nabla^2
\nabla^2 \sigma_s$.  Terms with an odd number of derivatives are explicitly
dissipative.  Examples are $\dot{\sigma}_s$, $\sigma_s \dot{\sigma}_s$, and
$\nabla^2 \dot{\sigma}_s$ \cite{CJK}.

Perhaps the closest analysis to ours is due to Rischke \cite{Rischke}.  The
differences may be summarized thusly:  First, Rischke used the influence
functional method \cite{IF}, which is closely related to the closed-time-path
method \cite{CTP}, for deriving the equations of motion of the classical field.
We use linear response theory.  Since the former techniques ultimately rely 
on a perturbative expansion in terms of $\sigma_s$ anyway, one might as
well employ linear response theory to begin with.  Linear response theory is
quicker, easier to use, and more intuitive.  Second, we write down the field
equation for $\sigma_s$, which is the deviation of the scalar condensate from
its equilibrium value $v$.  Rischke writes down the equation of motion for
${\bar \sigma} = v + \sigma_s$.  Of course these approaches are equivalent.
Third, Rischke's response functions are valid for free fields only.  For
example, a response function from Eqs. (\ref{delsig}--\ref{delpi}) has
the form
$\langle\left[\phi^2({\bf x}',t'), \phi^2({\bf x},t) \right] \rangle_{eq}$.
For free
fields this may be written $2[D_>^2(x'-x) - D_<^2(x'-x)]$ where the $D$ are
Wightmann functions/propagators and the subscript indicates whether $t'$ is
greater than or less than $t$.  Using our approach it is clear that these
response functions should be evaluated in the fully interacting ensemble (but
unperturbed by $\sigma_s$).  Taken together with the second difference this
constitutes a major improvement over Rischke's analysis.  This will be discussed
in more detail in the next section.  Fourth, Rischke employed a particular 
coarse graining technique by separating soft and hard modes according to the 
magnitude of the momentum.  We have left the coarse graining technique open.  
Finally, with a view towards the formation of disoriented chiral condensates, or 
DCC, Rischke allowed for slow classical components of the pion field.  We have 
not done so here, mainly to keep the analysis as simple and direct as possible,
although it could easily worked out in the same way as the sigma field.

\section{Estimating the Response Functions}

One's first inclination might be to evaluate the response functions using free
fields.  Suppose, for example, that $\sigma_s$ is so slowly varying in space and
time that to first approximation it can be taken outside the integration over
${\bf y}$ and $t'$.  Then it is a simple exercise to show that for a free field
$\phi$ with mass $m$ and energy E one gets
\begin{equation}
\int_0^t dt' \int d^3x' \left\langle \left[\phi^2({\bf x}',t'),
\phi^2({\bf x},t)
\right] \right\rangle_{eq} = i \int_0^t dt' \int \frac{d^3p}{(2\pi)^3}
\frac{1}{E^2}
\Bigl( 2 n_B(E/T) + 1 \Bigr) \sin[2E(t-t')]\;.
\end{equation}
The temperature-independent piece is a vacuum contribution and
may be dropped for the present discussion.  In the case that $m=0$ the momentum
integral is done trivially with the following result:
\begin{displaymath}
\frac{i}{4\pi^2} \int_0^t ds \left[ \frac{2\pi T}{\tanh(2\pi T s)} - \frac{1}{s}
\right] \, .
\end{displaymath}
In the limit that $t$ becomes large compared to $1/2\pi T$ this approaches the
asymptotic value $iTt/2\pi$.  This corresponds to the first term of a
Taylor expansion of $\sigma_s$ and further terms in the series bring in
powers of $(t-t')$ yielding
dissipative coefficients that grow as $t^2$, $t^3$ and
so on.  This is clearly unacceptable.  The origin of this problem can be traced
to the inadequacy of evaluating the response functions in the free field limit.
Indeed, these response functions are closely related to the shear and bulk
viscosities via the Kubo formulas which express those quantities in terms of
ensemble averages of commutators of the energy-momentum tensor density operator
at two different space-time points.  It is known that determination of the
viscosities requires summation of all ladder diagrams at a minimum \cite{Jeon1}.

An alternative is to make a harmonic approximation \cite{GM} instead of a Taylor
series expansion which, in this situation, can be formulated as:
\begin{displaymath}
\sigma_s(t-s) \approx \sigma_s(t) \cos(m_{\sigma}s) - \dot{\sigma}_s(t)
\sin(m_{\sigma}s)/ m_{\sigma} \, .
\end{displaymath}
For the applications we have in mind in this paper the deviation of the scalar
condensate from its equilibrium value is not expected to oscillate
significantly; rather, it is expected to decay exponentially to zero.
Therefore the harmonic approximation does not solve the problem.

Rather than attempt a sophisticated evaluation of the response functions in 
this paper we shall proceed to estimate them based on direct physical 
reasoning.  There are two obvious mechanisms for adding or removing sigma 
mesons from the condensate: (1) decay into two pions or the reverse process of 
formation via two pion annihilation, and (2) a meson from the thermal bath 
elastically scattering off a sigma meson and knocking it out of the condensate.  
The first of these processes is included in all analyses of the linear 
sigma model; the second of these processes has not been studied in the 
literature to our knowledge.  We shall evaluate them at the tree level.  This 
means that the parameters in the Lagrangian are to be fitted to experimental 
data at the tree level for consistency.  Let us examine each of these in turn.

The decay rate of a sigma meson into two pions in the sigma's rest frame is
\cite{decay}
\begin{equation}
\Gamma_{\sigma \rightarrow \pi\pi} = \frac{(N-1)}{8\pi}\frac{\lambda^2
v^2}{m_{\sigma}} = \frac{(N-1)}{16\pi} \lambda m_{\sigma}\;,
\end{equation}
when account is taken of the relation
$m_{\sigma}^2 = 2\lambda v^2$.  The decay
rate for a sigma meson at rest in the finite temperature system is Bose-enhanced
by a factor of $[1+n_B(m_{\sigma}/2T)]^2$, where $n_B$ is the Bose-Einstein
occupation number with the indicated argument of its exponential.  The rate for
two pions to form a sigma meson at rest is obtained from detailed balance by
multiplying $\Gamma_{\sigma \rightarrow \pi\pi}$ by $n_B^2(m_{\sigma}/2T)$.
Therefore the net rate is
\begin{eqnarray}
\Gamma_{\sigma\pi\pi} &=& \frac{(N-1)}{16\pi} \lambda m_{\sigma} \left[
(1+n_B)^2
- n_B^2 \right] = \frac{(N-1)}{16\pi} \lambda m_{\sigma} \left[ 1+2n_B \right]
\nonumber \\ &=&
\frac{(N-1)}{16\pi} \lambda m_{\sigma} \coth(m_{\sigma}/4T) \, .
\end{eqnarray}
(The argument of $n_B$ is the same as above.)  This leads to the dissipative
term $\Gamma_{\sigma\pi\pi} \dot{\sigma}_s$ in the equation of motion
\cite{remind}.  It agrees with eq. (80) of Rischke \cite{Rischke}.

Scattering of a thermal boson $b$ off a sigma meson with negligibly small
momentum and which is considered to be a component of the background field
$\sigma_s$ can be studied by evaluating the self-energy \cite{Shuryak,EK}.
\begin{eqnarray}
\Pi_{\sigma b} &=& - 4\pi \int \frac{d^3p}{(2\pi)^3} \,
n_B(E/T) \, \frac{\sqrt{s}}{E}
 \, f_{\sigma b}^{(\rm cm)}(s) \nonumber \\
&=& -\frac{2}{\pi} \int_{m_b}^{\infty} dE \,p\,
n_B(E/T) \sqrt{s} f_{\sigma b}^{(\rm cm)}(s)\;.
\end{eqnarray}
Here $E$ and $p$ are the energy and momentum of the boson $b$, $s=m_{\sigma}^2 +
m_b^2 + 2 m_{\sigma}E$, and $f_{\sigma b}$ is the forward scattering amplitude.
The normalization of the amplitude corresponds to the standard form of the
optical theorem
\begin{equation}
\sigma = \frac{4\pi}{q_{\rm cm}} {\rm Im} f^{(\rm cm)}(s) \, ,
\end{equation}
where $q_{\rm cm}$ is the momentum in the cm frame. Because the cross section is
invariant under longitudinal boosts the scattering amplitude transforms as
follows:
\begin{equation}
m_{\sigma} f_{\sigma b}^{(\sigma \,{\rm rest\,frame})} = \sqrt{s}
 f_{\sigma b}^{(\rm cm)} \, .
\end{equation}
The imaginary part of the self-energy
\begin{equation}
{\rm Im} \Pi_{\sigma b} = -\frac{m_{\sigma}}{2\pi^2} \int_{m_b}^{\infty} dE
\,p^2\, n_B(E/T) \sigma_{\sigma b}(s)
\end{equation}
determines the rate at which the field decays \cite{remind}:
\begin{equation}
\Gamma_{\sigma b} = - {\rm Im} \Pi_{\sigma b}/m_{\sigma} \, .
\end{equation}
The applicability of this expression is limited to those cases where
interference between sequential scatterings is negligible.

First consider pion scattering.  We will calculate to tree level only and
suppose that the parameters of the theory are adjusted to reproduce low energy
experimental data at this same level. The invariant amplitude is
\begin{equation}
{\cal M} = -2\lambda \left[ 1+m_{\sigma}^2 \left( \frac{1}{s} + \frac{1}{u} +
\frac{3}{t-m_{\sigma}^2} \right) \right] \, .
\end{equation}
The $s$, $t$ and $u$ are standard Mandelstam variables satisfying $s+t+u =
2m_{\sigma}^2$.  Note that the forward scattering amplitude evaluated at
threshold, ${\cal M}(s=m_{\sigma}^2,t=0)$, vanishes in accordance with Adler's
Theorem \cite{Adler}.  The differential cross section in the center-of-momentum
frame is
\begin{equation}
\frac{d\sigma}{d\Omega_{\rm cm}} = \frac{|{\cal M}|^2}{64 \pi^2 s} \, .
\end{equation}
The total cross section is given by
\begin{eqnarray}
\frac{4\pi s}{\lambda^2} \, \sigma &=& \frac{(s'+1)^2}{s'^2} +\frac{s'}{2-s'}
+\frac{9s'}{s'^2 -s' +1} \nonumber \\
&-& \frac{2(s'^2-3s'-1)}{(s'-1)^3} \ln\left[s'(2-s')\right] -\frac{6(s'^2-s'-
1)}{(s'-1)^3} \ln\left[ \frac{s'^2-s'+1}{s'} \right]\;,
\end{eqnarray}
where $s$ has been scaled by the sigma mass: $s' = s/m_{\sigma}^2$.  The total
cross section has a branch point and a pole at $s=2m_{\sigma}^2$
due to the $u$ channel exchange of a sigma meson.
Going beyond the tree level is necessary to incorporate the
finite width of the sigma meson and regulate the singularity.  Indeed, the tree
approximation is not reliable at high energy.  For example, it is well known
that unitarity is violated in $\pi\pi$ scattering at energies of order of one to
two times $m_{\sigma}$ \cite{pipi}.  Therefore we are only allowed to use the
low energy limit, which is quite acceptable when $T\ll m_{\sigma}$.  In this
limit
\begin{equation}
\sigma = \frac{112}{3\pi} \lambda^2 \frac{p^4}{m_{\sigma}^6} \, ,
\end{equation}
where $p$ is the pion momentum in the sigma rest frame.  Notice the workings of
Adler's Theorem here: According to that theorem the forward scattering amplitude
$f$ evaluated in the rest frame of any target particle must vanish like $p^2$ as
$p \rightarrow 0$.  Notice also that the total cross section cannot be obtained
from the imaginary part of $f$ because we are essentially using a Born
approximation.  In the limit that $T \rightarrow T_c$ we have the opposite
situation where $m_{\sigma}\ll T$.  Then the three-point vertices do not
contribute (${\cal M}=-2\lambda$ in place of (29)) and so use of
\begin{equation}
\sigma = \frac{\lambda^2}{4\pi s}
\end{equation}
may be considered more appropriate.

The contribution to the imaginary part of the self-energy is readily determined
in these two limits.  At low energy/temperature
\begin{equation}
{\rm Im} \Pi_{\sigma \pi} = -\frac{13,440}{\pi^3} \zeta(7) \lambda^2
\frac{T^7}{m_{\sigma}^5}\;,
\end{equation}
and at high energy/temperature
\begin{equation}
{\rm Im} \Pi_{\sigma \pi} = - \frac{\lambda^2 T^2}{96\pi} \, .
\end{equation}
These are the contributions from a single pion and must be multiplied by $N-1$
to factor in all pions.  For the accuracy required in this paper we can
construct a Pad\'e approximant to represent these results.
\begin{equation}
{\rm Im} \Pi_{\sigma \pi} \approx -(N-1)\frac{\lambda^2 T^2}{96\pi}
\:\frac{T^5}{[T^5 + (m_{\sigma}/10.568)^5]}\;.
\end{equation}
Notice that this contribution to $\Gamma$ diverges as $T \rightarrow T_c$ on
account of division by $m_{\sigma}$.  This ought to come as no surprise since
one is approaching a critical point where certain fluctuations diverge.

The above analysis may be repeated for a sigma meson knocking another out of the
condensate.  The invariant amplitude for $\sigma \sigma \rightarrow \sigma
\sigma$ is
\begin{equation}
{\cal M} = -6\lambda \left[ 1+ 3m_{\sigma}^2 \left( \frac{1}{s-m_{\sigma}^2} +
\frac{1}{u-m_{\sigma}^2} + \frac{1}{t-m_{\sigma}^2} \right) \right] \, ,
\end{equation}
reflecting the symmetry in the $s, t, u$ channels.  The total cross section
is
\begin{equation}
\frac{8\pi s}{9\lambda^2} \, \sigma = \left( \frac{s'+2}{s'-1}\right)^2 +
\frac{18}{s'-3} -\frac{12(s'^2 -3s' -1)}{(s'-4)(s'-2)(s'-1)} \ln(s'-3) \, .
\end{equation}
This expression has no singularities because $s' \ge 4$. The low energy limit is
\begin{equation}
\sigma = \frac{9}{2\pi} \frac{\lambda^2}{m_{\sigma}^2}\;,
\end{equation}
which gives rise to the low temperature limit
\begin{equation}
{\rm Im} \Pi_{\sigma \sigma} = - \frac{9}{2\pi^3} \lambda^2 T^2 {\rm e}^{-
m_{\sigma}/T} \, .
\end{equation}
In the high energy limit only the four-point vertex contributes, giving
\begin{equation}
\sigma = \frac{9}{8\pi} \frac{\lambda^2}{s}\;,
\end{equation}
which gives rise to the high temperature limit
\begin{equation}
{\rm Im} \Pi_{\sigma \sigma} = - \frac{3}{64\pi} \lambda^2 T^2 \, .
\end{equation}
This expression is identical to that calculated by Jeon \cite{Jeon1} and Rischke
\cite{Rischke} in a symmetric $\phi^4$ model to
which it can be compared.  The two limits can be combined in a Pad\'e
approximant as
\begin{equation}
{\rm Im} \Pi_{\sigma \sigma} \approx - \frac{9}{2\pi} \:
\frac{\lambda^2 T^2}{[96 +
\pi^2 \left({\rm e}^{m_{\sigma}/T} -1 \right)]} \;,
\end{equation}
which is useful for numerical computations.

The total rate is obtained by addition of all components, namely:
\begin{equation}
\Gamma = \Gamma_{\sigma\pi\pi} + \Gamma_{\sigma\pi} + \Gamma_{\sigma\sigma} \, ,
\end{equation}
where
\begin{equation}
\Gamma_{\sigma\pi} = - {\rm Im} \Pi_{\sigma\pi}/m_{\sigma}\;,
\end{equation}
and
\begin{equation}
\Gamma_{\sigma\sigma} = - {\rm Im} \Pi_{\sigma\sigma}/m_{\sigma} \, .
\end{equation}
Other works, such as \cite{Rischke}, do not include the latter two scattering
contributions arguing that they are of order $\lambda^2$, while
$\Gamma_{\sigma\pi\pi}$ is of order $\lambda$. While this is
correct at low $T$ when $m_\sigma$ is large, it becomes questionable near to
the critical temperature where $m_\sigma$ is small (see the discussion
of Fig. 1 below).

It should be noted that the scattering contributions diverge at the critical
temperature on account of division by the vanishing sigma mass.  This may be a
signal of the breakdown of the use of tree-level scattering amplitudes and
requires further investigation.

\section{Solution in an Expanding Fireball}

In this section we shall study solutions to the coarse-grained field equation in
several limits.  First, suppose that the volume and temperature are fixed in
time but that the system is slightly out of equilibrium in the sense that
$\sigma_s \neq 0$.  The field equation is then
\begin{equation}
\ddot{\sigma}_s + \Gamma(T) \dot{\sigma}_s + m_{\sigma}^2(T) \sigma_s =
 -\lambda
\left( 3v(T)\sigma_s^2 + \sigma_s^3 \right) \, .
\end{equation}
where $\Gamma$ is the sum of decay and scattering terms as given in the previous
section.  When $|\sigma_s|\ll v$ this equation can be linearized.  It
is equivalent to a simple damped harmonic oscillator.  The system is underdamped
if $m_{\sigma} > \Gamma/2$ and overdamped if $m_{\sigma} < \Gamma/2$.  We
choose $\lambda = 18$ so that the sigma mass in vacuum is $6 f_{\pi}$,
corresponding to the s-wave resonance observed in the $\pi \pi$ channel.  In
figure 1 we plot $m_{\sigma}$ and the individual contributions to $\Gamma$ as
functions of temperature $T$.  The field is overdamped when $T > 0.8T_c$.

Next consider the expansion of the system created in a high energy nuclear
collision.  If the beam energy is high enough it will form a quark-gluon plasma
with temperature greater than $T_c$.  This ``fireball'' will expand and cool,
eventually reaching $T_c$.  At this moment, say at time $t = t_c$, the initial
conditions for the coarse-grained field must be specified.
We will assume, for sake of illustration, that the system is locally
uniform so that spatial gradients are unimportant. Depending on whether
the system is expanding spherically or only longitudinally along the
beam axis ($D$ = 3 or 1, respectively) we obtain the modified equation
of motion
\begin{equation}
\ddot{\sigma}_s + \ddot{v} + \frac{D}{t} (\dot{\sigma}_s + \dot{v}) + 
\Gamma(T)\dot{\sigma}_s + m_{\sigma}^2(T) \sigma_s =
-\lambda \left( 3v(T)\sigma_s^2 + \sigma_s^3 \right) \, .
\end{equation}
In this situation $t$ is really the local, or proper, time and the term
proportional to $D/t$ may be thought to arise from either the d`Alembertian or
from a volume dilution term $[\dot{V}(t)/V(t)] \dot{\sigma}_s$
\cite{Gavin,Randrup} analogous to the Hubble expansion \cite{early}.
Perhaps the easiest way to obtain this equation is to derive the equation of 
motion for the total condensate $\bar{\sigma} = v + \sigma_s$ and then make the 
substitution.  Most authors actually solve the equation of motion for 
$\bar{\sigma}$, but this is a matter of taste.  Note that $v(T(t))$ is the 
instantaneous value of the equilibrium condensate and that is why no potential 
for it appears above.

In addition to the coarse-grained field equation one needs to know how the local
temperature evolves with time.  It must be assumed that it changes slowly enough
so that local equilibrium can be maintained by the rapidly fluctuating pion and
sigma fields.  In principle one ought to solve the equation
\begin{equation}
de/dt = -Dw/t
\end{equation}
where $e$ is the energy density, $w=e+P$ is the enthalpy and $P$ is the
pressure.  Rather than working out a detailed description of the equation of
state, which is really dominated by the rapidly fluctuating thermal fields, we
simply assume a free massless gas of pions where the pressure is proportional to
$T^4$.  The sigma meson has a mass small compared to $T$ only very near the
critical temperature, and its effect is therefore generally unimportant.  As a
consequence, the temperature falls with time according to the law
\begin{equation}
T(t) = T_c \left(\frac{t_c}{t}\right)^{D/3} \, .
\end{equation}
A reasonable numerical value for $t_c$ is 3 fm/c 
\cite{first1,first2,fast,BG}. The back reaction of $\sigma_s$ on the time 
evolution of the temperature is thereby neglected.  This is a quite reasonable 
approximation because very little of the total energy resides in the field 
$\sigma_s$.

The equation of motion may be solved by numerically evolving an
analytic solution in the neighborhood of $t_c$. As $t\rightarrow t_c$
the equilibrium condensate $v\rightarrow0$, however Eqs. (16) and (50)
indicate that the $\ddot{v}$ and $\dot{v}$ are singular, as are the
scattering contributions to the width $\Gamma$. The analytic behavior of
$\sigma_s$ as $t\rightarrow t_c$ is uniquely determined by requiring that the
derivatives of $\sigma_s$ exactly cancel these singularities, while
$\sigma_s\rightarrow0$.  The result is displayed in figure 2 for $D=1$ and in 
figure 3 for $D=3$.  The deviation from equilibrium $\sigma_s$ is maximal less 
than 1/2 fm/c after the critical temperature is passed, and it dies away with a 
time scale of about 2 fm/c. There is a hint of oscillatory motion in the 
solutions, but basically they are overdamped.

\section{Conclusion}

In this paper we have studied the dynamical evolution of the scalar
condensate in the $O(N)$ linear sigma model in out-of-equilibrium
situations.  Our method is based on the equation of motion.  Dissipation
arises because of the response of the correlation functions of the fast modes
to the slow modes of the fields.  This is treated with standard linear
response theory.  These response functions should be computed
exactly and used in the resulting dissipative, coarse-grained equation
of motion.  However, such explicit computations are generally not possible to
do. Therefore, we identified the physical mechanisms responsible
for the dissipation and estimated the corresponding response functions
based on them.  These mechanisms include the decay of sigma mesons in the
condensate, and the knockout of sigma mesons in the condensate due to
scattering with thermal sigma mesons and pions.  To our knowledge, the
latter physical mechanisms have not been studied before.

We then studied the dynamical evolution of the condensate in heavy ion
collisions, after the phase transition from quark-gluon plasma to hadrons,
and allowing for either one or three dimensional expansion of the hot
matter. These showed that thermal equilibrium was reestablished rather rapidly, 
with a time constant of order 2 fm/c. Clearly, much more could be studied along 
these same lines, including the formation and fate of disoriented chiral 
condensates (DCC).

The method we used in this paper is very general, and may be applied to
other theories, including nuclear matter, QCD, and electroweak theory.
Such work is underway.

\section*{Acknowledgements}

J.K. thanks the Institute of Technology at the University of Minnesota for
granting a single quarter leave in the spring of 1999 and the Theory Division 
at CERN for hospitality and support during that time.  S.J. was supported by
the Director, Office of Energy Research, Office of High Energy and Nuclear 
Physics, Division of Nuclear Physics, and by the Office of Basic Energy 
Sciences, Division of Nuclear Sciences, of the U.S.  Department of Energy
under Contract No. DE-AC03-76SF00098. This work was also supported by
the Department of Energy under grant DE-FG02-87ER40328,
by the NSF under travel grant INT-9602108 and by the
Norwegian Research Council.


\section*{Figures}

\begin{figure}[htb]
\centerline{\psfig{figure=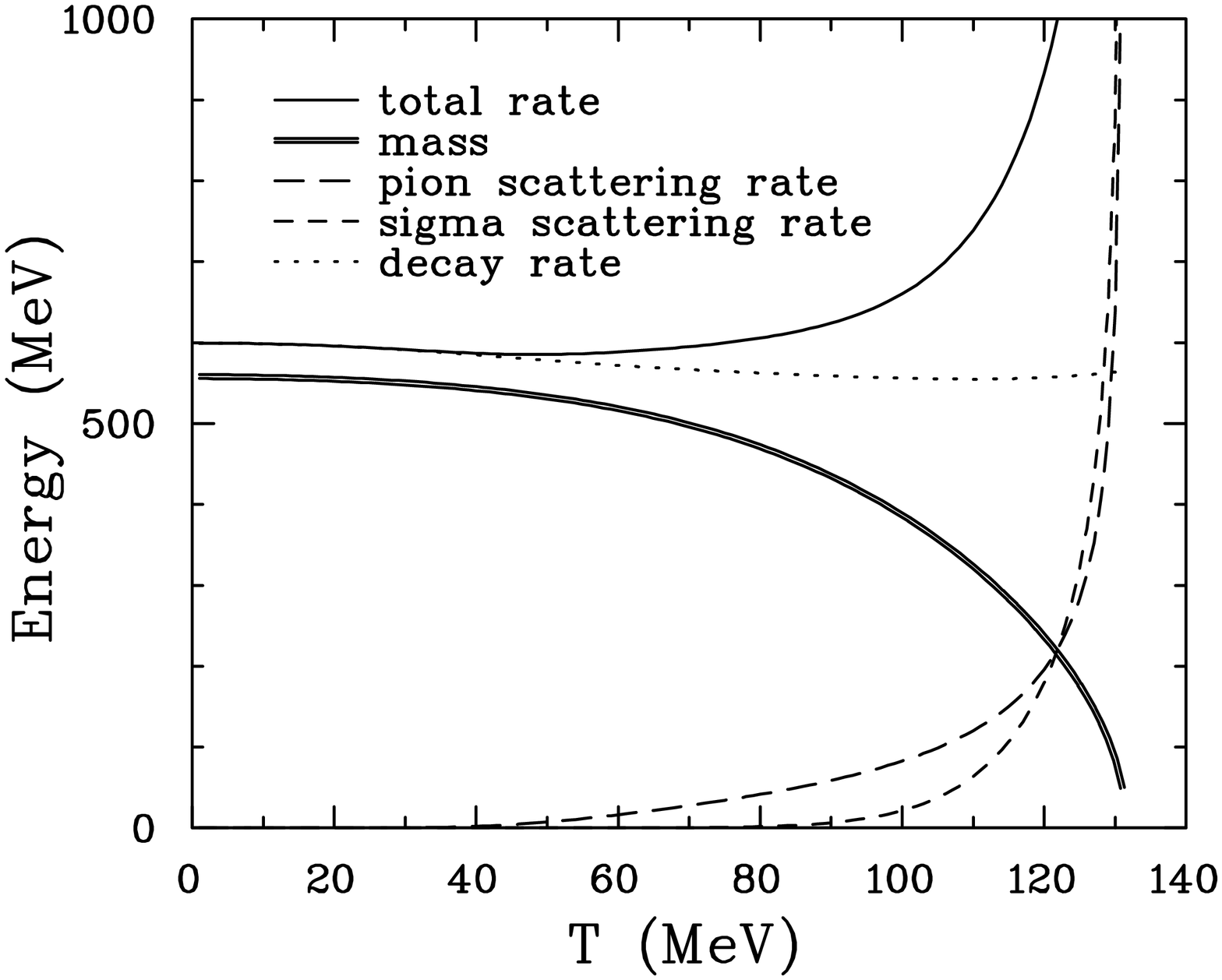,width=15cm,angle=90}}
\caption{The sigma mass and the decay and scattering contributions 
to the width
of the sigma meson as functions of temperature from 0 to $T_c$.}
\end{figure}

\begin{figure}[htb]
\centerline{\psfig{figure=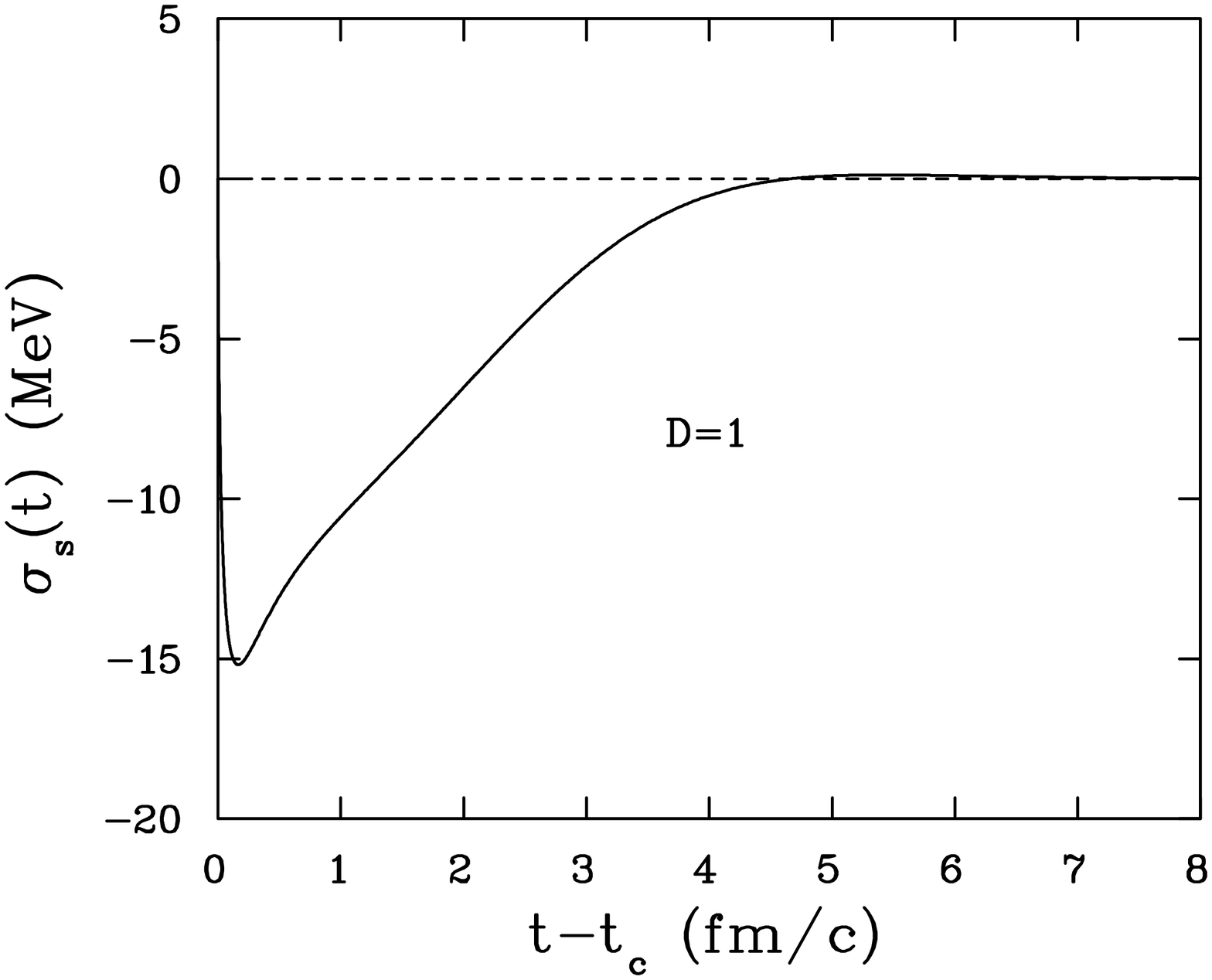,width=15cm,angle=90}}
\caption{The temporal evolution of the deviation of the scalar condensate
from its equilibrium value $v$ in units of MeV for a one dimensional expansion 
of hot matter produced in a high energy nuclear collision.}
\end{figure}

\begin{figure}[htb]
\centerline{\psfig{figure=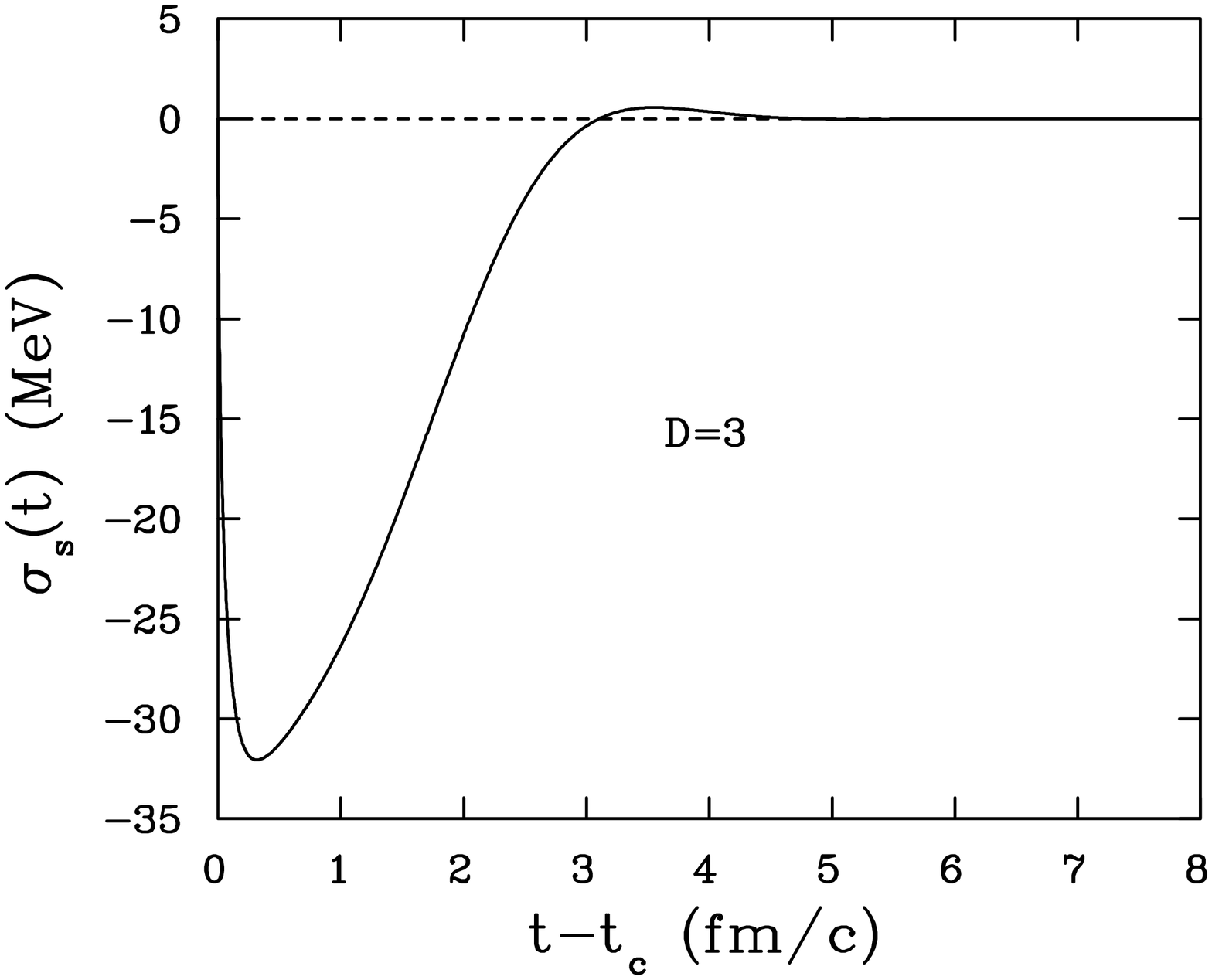,width=15cm,angle=90}}
\caption{The temporal evolution of the deviation of the scalar condensate
from its equilibrium value $v$ in units of MeV for a three 
dimensional expansion 
of hot matter produced in a high energy nuclear collision.}
\end{figure}

\end{document}